# The Liouville Dynamics of the q-Deformed 1-D Classical Harmonic Oscillator

[1]A. S. Mahmood and M. A. Z. Habeeb

Department of Physics, College of Science, Al-Nahrain University, Baghdad-Iraq.

[1]asm@sc.nahrainuniv.edu.iq

**Abstract**

The Liouville equation for the q-deformed 1-D classical harmonic oscillator is derived for two definitions of q-deformation. This derivation is achieved by using two different representations for the q-deformed Hamiltonian of this oscillator corresponding to undeformed and deformed phase spaces. The resulting Liouville equation is solved by using the method of characteristics in order to obtain the classical probability distribution function for this system. The 2-D and 3-D behaviors of this function are then investigated using a computer visualization method. The results are compared with those for the classical anharmonic oscillator. This comparison reveals that there are some similarities between these two models, where the results for the q-deformed oscillator exhibit similar whorl shapes that evolve with time as for the anharmonic oscillator. It is concluded that studying the Liouville dynamics gives more details about the physical nature of q-deformation than using the equation of motion method. It is also concluded that this result could have reflections on the interpretation of the quantized version of this q-deformed oscillator.

Keywords: Classical q-Deformed Oscillator, Liouville Equation, Method of Characteristics,

## Introduction

The notion of deformation is inherent in physics, where quantum mechanics can be considered as a deformation of Newtonian mechanics with deformation parameter $\hbar$ and, hence, in the limit $\hbar \to 0$ quantum mechanics reduces to Newtonian mechanics. Similarly, special relativity is the deformation of Newtonian mechanics with deformation parameter $v/c$ such that in the limit $v/c \to 0$ reduces to Newtonian mechanics [1]. From these physical examples, one can give a general mathematical definition for q-deformation in terms of a q-analog theorem [2, 3] where the identity expression is a generalization involving a new parameter that returns the original theorem in the limit $q \to 1$ [2, 3].

The q-deformed harmonic oscillator was introduced firstly in connection with studying quantum groups [4], where one can consider it as a deformation of the standard quantum harmonic oscillator. There are also different approaches to introduce q-deformation for the harmonic oscillator on the classical level. One approach, is the q-deformation of the Poisson bracket via the Jackson derivative [5]. Another approach, is the q-deformation of the Lagrangian of the harmonic oscillator [6]. Also, there is the possibility of q-deforming the action integral to obtain the q-deformed equation of motion [7]. On the physical interpretation side, the problem of q-deformation is still an open problem. This is true on both the classical and quantum mechanical levels. As a first step in tackling, this interpretation problem, we deal in this work with the classical q-deformed harmonic oscillator as a model for applying our methodology. Previously, some attempts in this direction can be found in the work of Man'ko [8] based on the equation of motion for this oscillator. In contrast, our methodology is based on deriving the Liouville equation for this q-deformed oscillator. We believe that this gives more details about the physical behavior of a system in its phase space compared with the equation of motion method which has known limitations.

It is also considered as a necessary first step before dealing with the problem of the quantized version of this oscillator [9].

The rest of the paper is organized as follows. First, the q-deformed classical harmonic oscillator this section is discussed where its Hamiltonian is introduced for both types of q-deformation. Then, the equations of motion and the Liouville equations are derived. Finally, the solutions for these Liouville equations are obtained by using the method of characteristics, then used to simulate the behavior in two and three dimensions and the conclusions are presented.

**The q-Deformed Classical Harmonic Oscillator**

The Hamiltonian of the 1-D simple harmonic oscillator (SHO) with mass $m$ and angular frequency $\omega$ is defined as:

$$H = \frac{p^2}{2m} + \frac{m\omega^2 q^2}{2} \qquad (1)$$

Let us also define:

$$\alpha = \sqrt{\frac{m\omega}{2\hbar}}\, q + \frac{ip}{\sqrt{2\hbar m\omega}} \qquad (2)$$

$$\alpha^* = \sqrt{\frac{m\omega}{2\hbar}}\, q - \frac{ip}{\sqrt{2\hbar m\omega}} \qquad (3)$$

where the appearance of $\hbar$ in these equations is to provide a convenient scaling for various physical quantities [10]. In terms of these new complex coordinates $(\alpha, \alpha^*)$, the canonical coordinates $(q, p)$ can be written as [8, 10, 11]:

$$q = \sqrt{\frac{\hbar}{2m\omega}}\, (\alpha + \alpha^*) \qquad (4)$$

$$p = -i\sqrt{\frac{\hbar m\omega}{2}}\, (\alpha - \alpha^*) \qquad (5)$$

Then, substituting eqns. (4) and (5) into eqn. (1) one obtains [8, 10]:

$$H(\alpha, \alpha^*) = \hbar\omega\alpha\alpha^* \qquad (6)$$

The two complex variables $\alpha$ and $\alpha^*$ can be considered as two independent coordinates in a complex phase space [11].

The concept of q-deformation can be introduced into this 1-D SHO by transforming to q-deformed coordinates $\alpha_q$ and $\alpha_q^*$ by a non-linear transformation as [8,13]:

$$\alpha_q = f(\alpha, \alpha^*)\, \alpha \qquad (7)$$

$$\alpha_q^* = f^*(\alpha, \alpha^*)\, \alpha^* \qquad (8)$$

where $f = f(\alpha, \alpha^*)$ is a non-negative function of the two independent complex variables $\alpha$ and $\alpha^*$.

Generally, the function $f(\alpha, \alpha^*)$ has the following three forms corresponding to the deformation types shown in Table (1):

**Table (1): Forms of the function $f(\alpha, \alpha^*)$ and their associated types of deformation.**

| $f(\alpha, \alpha^*)$ | type of deformation |
|---|---|
| 1 | undeformed case |
| $\sqrt{\dfrac{[\alpha\alpha^*]_q}{\alpha\alpha^*}}$ | q–deformed case |
| otherwise | general f–deformed case |

where $[\ ]_q$ is a q-number [8,13,14] defined as either:

$$[\alpha\alpha^*]_q = \frac{q^{\alpha\alpha^*} - q^{-\alpha\alpha^*}}{q - q^{-1}}$$

hence,

$$\sqrt{\frac{[\alpha\alpha^*]_q}{\alpha\alpha^*}} = \sqrt{\frac{\sinh(\lambda\alpha\alpha^*)}{\alpha\alpha^* \sinh(\lambda)}} \quad (9)$$

or

$$[\alpha\alpha^*]_q = \frac{q^{\alpha\alpha^*} - 1}{q - 1}$$

hence,

$$\sqrt{\frac{[\alpha\alpha^*]_q}{\alpha\alpha^*}} = \sqrt{\frac{e^{\lambda\alpha\alpha^*} - 1}{\alpha\alpha^* (e^\lambda - 1)}} \quad (10)$$

where,

$$q = e^\lambda \quad (11)$$

The parameter $q$ represents a real q-deformation parameter in the range of values $0 < q < 1$ while $\lambda$ represents the nonlinearity parameter in the range of values $-\infty < \lambda < 0$.

(i) The Classical Hamiltonian in $\alpha_q$ –Representation

The classical Hamiltonian is defined in terms of the $(\alpha_q, \alpha_q^*)$ coordinates as given by Man'ko [8, 13] as

$$\mathcal{H}_q(\alpha_q, \alpha_q^*) = \hbar\omega \alpha_q \alpha_q^* \quad (12)$$

(ii) The Classical Hamiltonian in α–Representation

Using the definition of the function $f$ given by eqns. (9) and (10) in the non-linear transformation eqns. (7) and (8), then substituting the results into eqn. (12), we obtain:

$$\mathbb{H}_q(\alpha, \alpha^*) = \hbar\omega f f^* \alpha\alpha^* =$$

$$\hbar\omega \left\{ \frac{\sinh(\lambda\alpha\alpha^*)}{\sinh(\lambda)} \right\} \quad (13a)$$

and,

$$\mathbb{H}_q(\alpha, \alpha^*) = \hbar\omega f f^* \alpha\alpha^* =$$

$$\hbar\omega \left\{ \frac{e^{\lambda\alpha\alpha^*} - 1}{(e^\lambda - 1)} \right\} \quad (13b)$$

respectively, where $\mathbb{H}_q(\alpha, \alpha^*)$ represents the Hamiltonian of the q-deformed oscillator defined in terms of the complex undeformed $(\alpha, \alpha^*)$ coordinates.

Eqn. (13a) was previously obtained by Man'ko [8, 13].

**The Equation of Motion**

The equation of motion for the q–deformed harmonic oscillator can be obtained in the $\alpha$-representation and $\alpha_q$-representation as:

(i) $\alpha$ –Representation:

The equation of motion in this representation is:

$$\dot{\alpha}(t) = \{\alpha, \mathbb{H}_q(\alpha, \alpha^*)\}_{q,p} \quad (14)$$

where the notation $\{\,,\,\}_{q,p}$ represents the Poisson bracket with respect to the canonical coordinates $(q, p)$, where the subscripts $(q, p)$ will be dropped from now and on.

But since,

$$\{\alpha, \mathbb{H}_q(\alpha, \alpha^*)\} =$$

$$\{\alpha, \alpha^*\} \cdot \{\alpha, \mathbb{H}_q(\alpha, \alpha^*)\}_{\alpha,\alpha^*}$$

(see Appendix-A) $\quad (15)$

then simplifying $\{\alpha, \mathbb{H}_q(\alpha,\alpha^*)\}_{\alpha,\alpha^*}$ by using the fact that $\alpha$ and $\alpha^*$ are two independent complex variables, the result becomes

$$\{\alpha, \mathbb{H}_q(\alpha,\alpha^*)\}_{\alpha,\alpha^*} = \left(\frac{\partial \mathbb{H}_q(\alpha,\alpha^*)}{\partial \alpha^*}\right)_\alpha \quad (16)$$

and substituting $\mathbb{H}_q(\alpha,\alpha^*)$ from eqns. (13) into eqns. (16), these equations become:

$$\{\alpha, \mathbb{H}_q(\alpha,\alpha^*)\}_{\alpha,\alpha^*} = \hbar\omega \left\{\frac{\lambda \cosh(\lambda|\alpha|^2)}{\sinh(\lambda)}\right\}\alpha \quad (17a)$$

and,

$$\{\alpha, \mathbb{H}_q(\alpha,\alpha^*)\}_{\alpha,\alpha^*} = \hbar\omega \left\{\frac{\lambda e^{\lambda|\alpha|^2}}{(e^\lambda - 1)}\right\}\alpha \quad (17b)$$

Defining $\omega_q^{(\mu)}$

$$\omega_q^{(1)} = \omega \left\{\frac{\lambda \cosh(\lambda|\alpha|^2)}{\sinh(\lambda)}\right\} \quad (18a)$$

and,

$$\omega_q^{(2)} = \omega \left\{\frac{\lambda e^{\lambda|\alpha|^2}}{(e^\lambda - 1)}\right\} \quad (18b)$$

eqns. (17) take the form:

$$\{\alpha, \mathbb{H}_q(\alpha,\alpha^*)\}_{\alpha,\alpha^*} = \hbar \omega_q^{(\mu)} \alpha \quad ; \mu = 1, 2. \quad (19)$$

where $\omega_q^{(\mu)}$ can be considered as the frequency of the q–deformed classical harmonic oscillator in the $\alpha$–representation. Again, it is noticed that eqn. (18a) is the same as that introduced by Man'ko [8].

Also, it can be shown that [8, 10, 11]:
$$\{\alpha, \alpha^*\} = -\left(\frac{i}{\hbar}\right) \quad (20)$$

Then, substituting eqns. (19) and (20) into eqn. (15), and using the result in eqn. (14), yields

$$\dot{\alpha}(t) = -i\,\omega_q^{(\mu)} \alpha \quad ; \mu = 1, 2. \quad (21)$$

It is worth mentioning that, $|\alpha|^2$ represents a constant of the motion for the undeformed classical oscillator. This is based on the fact that the Poisson bracket for the undeformed Hamiltonian $\{\alpha\alpha^*, H(\alpha,\alpha^*)\}_{\alpha,\alpha^*} = 0$, where $H(\alpha,\alpha^*)$ is as given by eqn. (6). A similar result can be obtained for $\{\alpha\alpha^*, \mathbb{H}_q(\alpha,\alpha^*)\}_{\alpha,\alpha^*} = 0$ and $\{\alpha_q\alpha_q^*, \mathcal{H}_q(\alpha_q,\alpha_q^*)\}_{\alpha_q,\alpha_q^*} = 0$.

Eqn. (21) and its complex conjugate represent the equations of motion for the q-deformed classical harmonic oscillator in the $\alpha$-representation.

Solving these equations of motion gives the equations of trajectories for the q-deformed classical harmonic oscillator in the complex $(\alpha, \alpha^*)$ phase space:

$$\left.\begin{array}{l}\alpha(t) = \alpha(0)\,e^{-i\,\omega_q^{(\mu)} t} \\ \alpha^*(t) = \alpha^*(0)\,e^{i\,\omega_q^{(\mu)} t}\end{array}\right\} \quad ; \mu = 1, 2. \quad (22)$$

where, $\alpha(0)$ and $\alpha^*(0)$ are initial trajectory points at $t = 0$.

(ii) $\alpha_q$–Representation:

In this case, the equation of motion is given by:

$$\dot{\alpha}_q(t) = \{\alpha_q, \mathcal{H}_q(\alpha_q,\alpha_q^*)\} \quad (23)$$

where,

$$\{\alpha_q, \mathcal{H}_q(\alpha_q, \alpha_q^*)\} =$$
$$\{\alpha_q, \alpha_q^*\} \cdot \{\alpha_q, \mathcal{H}_q(\alpha_q, \alpha_q^*)\}_{\alpha_q, \alpha_q^*}$$
(see Appendix-A) (24)

Also,
$$\{\alpha_q, \mathcal{H}_q(\alpha_q, \alpha_q^*)\}_{\alpha_q, \alpha_q^*} = \left(\frac{\partial \mathcal{H}_q(\alpha_q, \alpha_q^*)}{\partial \alpha_q^*}\right)_{\alpha_q} \quad (25)$$

Substituting $\mathcal{H}_q(\alpha_q, \alpha_q^*)$ from eqn. (12) into eqn. (25), the result is:
$$\{\alpha_q, \mathcal{H}_q(\alpha_q, \alpha_q^*)\}_{\alpha_q, \alpha_q^*} = \hbar \omega \alpha_q \quad (26)$$

Also, since the Poisson bracket $\{\alpha_q, \alpha_q^*\} = \{\alpha, \alpha^*\} \cdot \{\alpha_q, \alpha_q^*\}_{\alpha, \alpha^*}$ for the q-deformed oscillator can be written as:

$$\{\alpha_q, \alpha_q^*\} = -\left(\frac{i}{\hbar}\right)\left\{\frac{\lambda \sqrt{1+|\alpha_q|^4 \sinh^2(\lambda)}}{\sinh(\lambda)}\right\}$$
(27a)

or,
$$\{\alpha_q, \alpha_q^*\} = -\left(\frac{i}{\hbar}\right)\left\{\frac{\lambda\left[1-|\alpha_q|^2(1-e^\lambda)\right]}{(e^\lambda - 1)}\right\}$$
(27b)

Then substituting the Poisson brackets $\{\alpha_q, \mathcal{H}_q(\alpha_q, \alpha_q^*)\}_{\alpha_q, \alpha_q^*}$ and $\{\alpha_q, \alpha_q^*\}$ from eqn. (26) and eqns. (27) into eqn. (24), and using the result in eqn. (23), one obtains:
$$\dot{\alpha}_q(t) = -i\omega_q^{(\mu)} \alpha_q \quad ; \mu = 3, 4. \quad (28)$$
where,
$$\omega_q^{(3)} = \omega \left\{\frac{\lambda \sqrt{1+|\alpha_q|^4 \sinh^2(\lambda)}}{\sinh(\lambda)}\right\} \quad (29a)$$

and,
$$\omega_q^{(4)} = \omega \left\{\frac{\lambda\left[1-|\alpha_q|^2(1-e^\lambda)\right]}{(e^\lambda - 1)}\right\} \quad (29b)$$

It is noticed that eqn. (29a) is the same as that introduced by Man'ko [8].

Eqn. (28) and its complex conjugate represent the equations of motion for the q-deformed classical harmonic oscillator in the $\alpha_q$-representation.

Again, solving these equations of motion, gives the equations of trajectories for the q-deformed classical harmonic oscillator in the complex $(\alpha_q, \alpha_q^*)$ phase space:

$$\left.\begin{array}{l} \alpha_q(t) = \alpha_q(0) e^{-i\omega_q t} \\ \alpha_q^*(t) = \alpha_q^*(0) e^{i\omega_q t} \end{array}\right\} \quad (30)$$

where, $\alpha_q(0)$ and $\alpha_q^*(0)$ are initial trajectory points at $t = 0$.

**The Liouville Equation**

In this section, the Liouville equation for the q-deformed classical harmonic oscillator in the two complex phase space representations is derived. In general, the Liouville equation for a Hamiltonian system described by a Hamiltonian H is given as [12]:

$$\frac{\partial \mathcal{A}}{\partial t} = \{H, \mathcal{A}\} \quad (31)$$

where $\mathcal{A}$ represents any dynamical variable.

(i) Liouville's Equation in the $\alpha$-Representation

In this case, and using the Hamiltonian $\mathbb{H}_q(\alpha, \alpha^*)$, the classical Liouville equation can be shown to be:

$$\frac{\partial \mathcal{P}_{CL}^{q}(\alpha,\alpha^*;t)}{\partial t} = \left\{\mathbb{H}_q(\alpha,\alpha^*), \mathcal{P}_{CL}^{q}(\alpha,\alpha^*;t)\right\} \quad (32)$$

where $\mathcal{P}_{CL}^{q}(\alpha,\alpha^*;t)$ represents the probability distribution function for the q-deformed classical oscillator in the $\alpha$-representation, where the subscript $CL$ indicates the classical system.

But since,

$$\left\{\mathbb{H}_q(\alpha,\alpha^*), \mathcal{P}_{CL}^{q}(\alpha,\alpha^*;t)\right\} =$$

$$\left\{\alpha,\alpha^*\right\} \cdot \left\{\mathbb{H}_q(\alpha,\alpha^*), \mathcal{P}_{CL}^{q}(\alpha,\alpha^*;t)\right\}_{\alpha,\alpha^*}$$

(see Appendix-B) (33)

then, using the definition of the Hamiltonian $\mathbb{H}_q(\alpha,\alpha^*)$ in the $\alpha$-representation as in eqns. (13), we obtain:

$$\left(\frac{\partial \mathbb{H}_q(\alpha,\alpha^*)}{\partial \alpha}\right)_{\alpha^*} = \hbar\omega\left(f^2 + 2\alpha f f_\alpha\right)\alpha^* \quad (34)$$

and,

$$\left(\frac{\partial \mathbb{H}_q(\alpha,\alpha^*)}{\partial \alpha^*}\right)_\alpha = \hbar\omega\left(f^2 + 2\alpha^* f f_{\alpha^*}\right)\alpha \quad (35)$$

Furthermore, the Poisson bracket $\left\{\mathbb{H}_q(\alpha,\alpha^*), \mathcal{P}_{CL}^{q}(\alpha,\alpha^*;t)\right\}_{\alpha,\alpha^*}$ can be written as:

$$\left\{\mathbb{H}_q(\alpha,\alpha^*), \mathcal{P}_{CL}^{q}(\alpha,\alpha^*;t)\right\}_{\alpha,\alpha^*} =$$

$$\left(\frac{\partial \mathbb{H}_q(\alpha,\alpha^*)}{\partial \alpha}\right)_{\alpha^*}\left(\frac{\partial \mathcal{P}_{CL}^{q}(\alpha,\alpha^*;t)}{\partial \alpha^*}\right)$$

$$-\left(\frac{\partial \mathbb{H}_q(\alpha,\alpha^*)}{\partial \alpha^*}\right)_\alpha\left(\frac{\partial \mathcal{P}_{CL}^{q}(\alpha,\alpha^*;t)}{\partial \alpha}\right) \quad (36)$$

Then, substituting eqns. (34) and (35) into eqn. (36), this Poisson bracket becomes:

$$\left\{\mathbb{H}_q(\alpha,\alpha^*), \mathcal{P}_{CL}^{q}(\alpha,\alpha^*;t)\right\}_{\alpha,\alpha^*} =$$

$$\hbar\omega\left(\left(f^2 + 2\alpha f f_\alpha\right)\alpha^*\frac{\partial}{\partial \alpha^*}\right.$$

$$\left. -\left(f^2 + 2\alpha^* f f_{\alpha^*}\right)\alpha\frac{\partial}{\partial \alpha}\right)\mathcal{P}_{CL}^{q}(\alpha,\alpha^*;t) \quad (37)$$

Substituting eqns. (37) and the Poisson bracket $\left\{\alpha,\alpha^*\right\}$ from eqn. (20) into eqn. (33), we can write:

$$\left\{\mathbb{H}_q(\alpha,\alpha^*), \mathcal{P}_{CL}^{q}(\alpha,\alpha^*;t)\right\} = -i\omega$$

$$\cdot \left(\left(f^2 + 2\alpha f f_\alpha\right)\alpha^*\frac{\partial}{\partial \alpha^*}\right.$$

$$\left. -\left(f^2 + 2\alpha^* f f_{\alpha^*}\right)\alpha\frac{\partial}{\partial \alpha}\right)\mathcal{P}_{CL}^{q}(\alpha,\alpha^*;t) \quad (38)$$

Using eqn. (38) in eqn. (32) and re-arranging, the result becomes:

$$\frac{\partial \mathcal{P}_{CL}^{q}(\alpha,\alpha^*;t)}{\partial t} =$$

$$-i\omega\left\{\left(f^2 + 2\alpha f f_\alpha\right)\alpha^*\frac{\partial}{\partial \alpha^*}\right.$$

$$\left. -\left(f^2 + 2\alpha^* f f_{\alpha^*}\right)\alpha\frac{\partial}{\partial \alpha}\right\}\mathcal{P}_{CL}^{q}(\alpha,\alpha^*;t)$$

(39)

But, since $\alpha f_\alpha = \alpha^* f_{\alpha^*}$ where

$$\alpha f_\alpha = \alpha^* f_{\alpha^*} =$$

$$\left(\frac{1}{2f}\right)\left\{\frac{\lambda\cosh(\lambda|\alpha|^2)}{\sinh(\lambda)} - f^2\right\} \quad (40a)$$

and

$$\alpha f_\alpha = \alpha^* f_{\alpha^*} =$$

$$\left(\frac{1}{2f}\right)\left\{\frac{\lambda e^{\lambda|\alpha|^2}}{(e^\lambda - 1)} - f^2\right\} \quad (40b)$$

then, eqn. (39) becomes:

$$\frac{\partial \mathcal{P}_{CL}^q(\alpha,\alpha^*;t)}{\partial t} =$$

$$-i\omega_q^{(\mu)}\left(\alpha^*\frac{\partial}{\partial \alpha^*} - \alpha\frac{\partial}{\partial \alpha}\right)\mathcal{P}_{CL}^q(\alpha,\alpha^*;t)$$

$$; \mu = 3,4. \quad (41)$$

where $\omega_q^{(\mu)}$ is given by eqns. (18).

Eqn. (41) represent the Liouville equation for the q-deformed classical harmonic oscillator in the $\alpha$-representation.

(ii) Liouville's Equation in the

$\alpha_q$–Representation

For this case, the Liouville equation is given as:

$$\frac{\partial P_{CL}^q(\alpha_q,\alpha_q^*;t)}{\partial t} =$$

$$\left\{\mathcal{H}_q(\alpha_q,\alpha_q^*), P_{CL}^q(\alpha_q,\alpha_q^*;t)\right\} \quad (42)$$

where $P_{CL}^q(\alpha_q,\alpha_q^*;t)$ represents the probability distribution function for the q-deformed classical harmonic oscillator in the $\alpha_q$-representation.

But since,

$$\left\{\mathcal{H}_q(\alpha_q,\alpha_q^*), P_{CL}^q(\alpha_q,\alpha_q^*;t)\right\} = \left\{\alpha_q,\alpha_q^*\right\}$$

$$\cdot\left\{\mathcal{H}_q(\alpha_q,\alpha_q^*), P_{CL}^q(\alpha_q,\alpha_q^*;t)\right\}_{\alpha_q,\alpha_q^*}$$

(see Appendix-B) (43)

and the Poisson bracket $\left\{\mathcal{H}_q(\alpha_q,\alpha_q^*), P_{CL}^q(\alpha_q,\alpha_q^*;t)\right\}_{\alpha_q,\alpha_q^*}$ is defined as:

$$\left\{\mathcal{H}_q(\alpha_q,\alpha_q^*), P_{CL}^q(\alpha_q,\alpha_q^*;t)\right\}_{\alpha_q,\alpha_q^*} =$$

$$\left(\frac{\partial \mathcal{H}_q(\alpha_q,\alpha_q^*)}{\partial \alpha_q}\right)_{\alpha_q^*}\left(\frac{\partial P_{CL}^q(\alpha_q,\alpha_q^*;t)}{\partial \alpha_q^*}\right)$$

$$-\left(\frac{\partial \mathcal{H}_q(\alpha_q,\alpha_q^*)}{\partial \alpha_q^*}\right)_{\alpha_q}\left(\frac{\partial P_{CL}^q(\alpha_q,\alpha_q^*;t)}{\partial \alpha_q}\right) \quad (44)$$

then, using the definition of the Hamiltonian $\mathcal{H}_q(\alpha_q,\alpha_q^*)$ in the $\alpha_q$-representation as in eqn. (12), gives:

$$\left(\frac{\partial \mathcal{H}_q(\alpha_q,\alpha_q^*)}{\partial \alpha_q}\right)_{\alpha_q^*} = \omega\alpha_q^* \quad (45)$$

and,

$$\left(\frac{\partial \mathcal{H}_q(\alpha_q,\alpha_q^*)}{\partial \alpha_q^*}\right)_{\alpha_q} = \omega\alpha_q \quad (46)$$

Substituting eqns. (45) and (46) into eqn. (44) and re-arranging, the result takes the form:

$$\left\{\mathcal{H}_q(\alpha_q,\alpha_q^*), P_{CL}^q(\alpha_q,\alpha_q^*;t)\right\}_{\alpha_q,\alpha_q^*} =$$

$$\omega\left(\alpha_q^*\frac{\partial}{\partial \alpha_q^*} - \alpha_q\frac{\partial}{\partial \alpha_q}\right)P_{CL}^q(\alpha_q,\alpha_q^*;t)$$

(47)

Now, substituting eqn. (47) and the Poisson bracket $\{\alpha_q,\alpha_q^*\}$ from eqns. (27) into eqn. (43), we obtain:

$$\left\{\mathcal{H}_q(\alpha_q,\alpha_q^*), P_{CL}^q(\alpha_q,\alpha_q^*;t)\right\} =$$

$$-i\omega_q^{(\mu)}\left(\alpha_q^*\frac{\partial}{\partial \alpha_q^*} - \alpha_q\frac{\partial}{\partial \alpha_q}\right)$$

$$\cdot P_{CL}^q(\alpha_q,\alpha_q^*;t) \quad (48)$$

Substituting eqn. (48) into eqn. (42), gives:

$$\frac{\partial P_{CL}^{q}(\alpha_q, \alpha_q^*; t)}{\partial t} = -i\omega_q^{(\mu)}$$

$$\cdot \left( \alpha_q^* \frac{\partial}{\partial \alpha_q^*} - \alpha_q \frac{\partial}{\partial \alpha_q} \right) P_{CL}^{q}(\alpha_q, \alpha_q^*; t)$$

$$; \mu = 3, 4. \quad (49)$$

where, $\omega_q$ is given by eqns. (29).

Eqn. (49) is the required Liouville equation for the q-deformed classical oscillator in the $\alpha_q$-representation.

It is obvious that in the limit $q \to 1$, the Liouville eqns. (41) and (49) for the q-deformed classical harmonic oscillator in the $\alpha$- and $\alpha_q$-representations reduce to the Liouville equation of the undeformed harmonic oscillator with angular frequency $\omega$:

$$\frac{\partial P_{CL}(\alpha, \alpha^*; t)}{\partial t} =$$

$$-i\omega \left( \alpha^* \frac{\partial}{\partial \alpha^*} - \alpha \frac{\partial}{\partial \alpha} \right) P_{CL}(\alpha, \alpha^*; t)$$

$$(50)$$

where $P_{CL}(\alpha, \alpha^*; t)$ represents the probability distribution function for this oscillator in the $\alpha$-representation as expected.

**Solutions of the Liouville Equations**

Solutions of the Liouville equations of the q-deformed harmonic oscillator, (i.e., eqns. (41) and (49)) can be obtained by using the method of characteristics [10] in the same manner as performed by Milburn [10] for the anharmonic oscillator. Hence, in the $\alpha$-representation, assuming the following initial solution at $t = 0$:

$$P_{CL}^{q}(\alpha, \alpha^*; 0) = e^{-|\alpha - \alpha(0)|^2} \quad (51)$$

then, by transforming to the rotating frame [10], the time-evolution for each point $(\alpha, \alpha^*)$ in complex phase space can be obtained by replacing $\alpha$ by $\alpha(t)$ given by [10]:

$$\alpha(t) = \alpha e^{-i\omega_q^{(\mu)} t} \quad (52)$$

Substituting eqn. (52) into eqn. (51), one obtains:

$$P_{CL}^{q}(\alpha, \alpha^*; t) = e^{-\left|\alpha e^{-i\omega_q^{(\mu)} t} - \alpha(0)\right|^2} \quad (53)$$

Similarly, in the $\alpha_q$-representation, we assume that:

$$P_{CL}^{q}(\alpha_q, \alpha_q^*; t) =$$

$$e^{-\left|\alpha_q e^{-i\omega_q^{(\mu)} t} - \alpha_q(0)\right|^2} \quad (54)$$

By direct substitution of the solutions given in eqns. (53) and (54) into the classical Liouville eqns. (41) and (49), one can verify that these solutions satisfy these equations.

**Computer Visualizations for the q-Deformed Classical Harmonic Oscillator**

The computer visualization method is utilized by Milburn [10] to investigate the time-evolution of the probability distribution function for the anharmonic oscillator in phase-space using the Husimi $Q$-function. This function represents a quasiprobability distribution function and it is widely used in the field of quantum optics. The $Q$-function is a normalizable positive function with values in the range $0 < Q < 1$. However, it is noted that the frequency of the anhrmonic oscillator treated by Milburn [10] is a function of $|\alpha|^2$. Similarly, for the case of the q-deformed oscillator the frequency is a function of $|\alpha|^2$ or $|\alpha_q|^2$ depending on the representation. This motivates using the same method used by Milburn [10] to investigate the behavior of the probability distribution functions $\mathcal{P}_{CL}^q(\alpha,\alpha^*;t)$ and $P_{CL}^q(\alpha_q,\alpha_q^*;t)$.

In the present work, a computer simulation program was written in Mathematica® to perform the computer visualizations. The initial values $q_o$ and $p_o$ were taken as $\frac{1}{\sqrt{2}}$, 0 measured in units of $\sqrt{\frac{2\hbar}{m\omega}}$ and $\sqrt{2\hbar m\omega}$ respectively and, hence, from eqn. (2) $\alpha(0) = 0.5$.

To verify the reliability of this simulation program, the same results that were obtained by Milburn [10], were found to be reproducible in the present work by applying the computer program to his problem [10]. The results obtained are illustrated in Fig. (1). Using the dimensionless time $\tau = \omega t$.

In a similar manner, the time evolution of the probability distribution function $\mathcal{P}_{CL}^q(\alpha,\alpha^*;t)$ can be represented in phase space through the behavior of a particular initial contour $|\alpha-\alpha(0)| = \frac{1}{2}$ centered at $\alpha(0)$ [10]. Each point on the initial contour will move according to eqns. (23), and the evolution of this initial contour within the time interval $0 \leq \tau \leq 2\pi$ in the phase space region $-1 \leq q \leq 1$ and $-1 \leq p \leq 1$ was followed. The results of such a procedure are depicted in Fig. (2) and Fig (3) where the time-evolution of the 2-D probability distribution is shown. The results for the time evolution of the probability distribution function $P_{CL}^q(\alpha_q,\alpha_q^*;t)$ have the same values despite the fact that they different expressions. For this reason, these results will not be presented here. These figures exhibit whorl shapes and can be contrasted with those obtained by Milburn [10] for the anharmonic oscillator as shown in Fig. (1), where it is clear that these whorl shapes again become finer as $t \to \infty$.

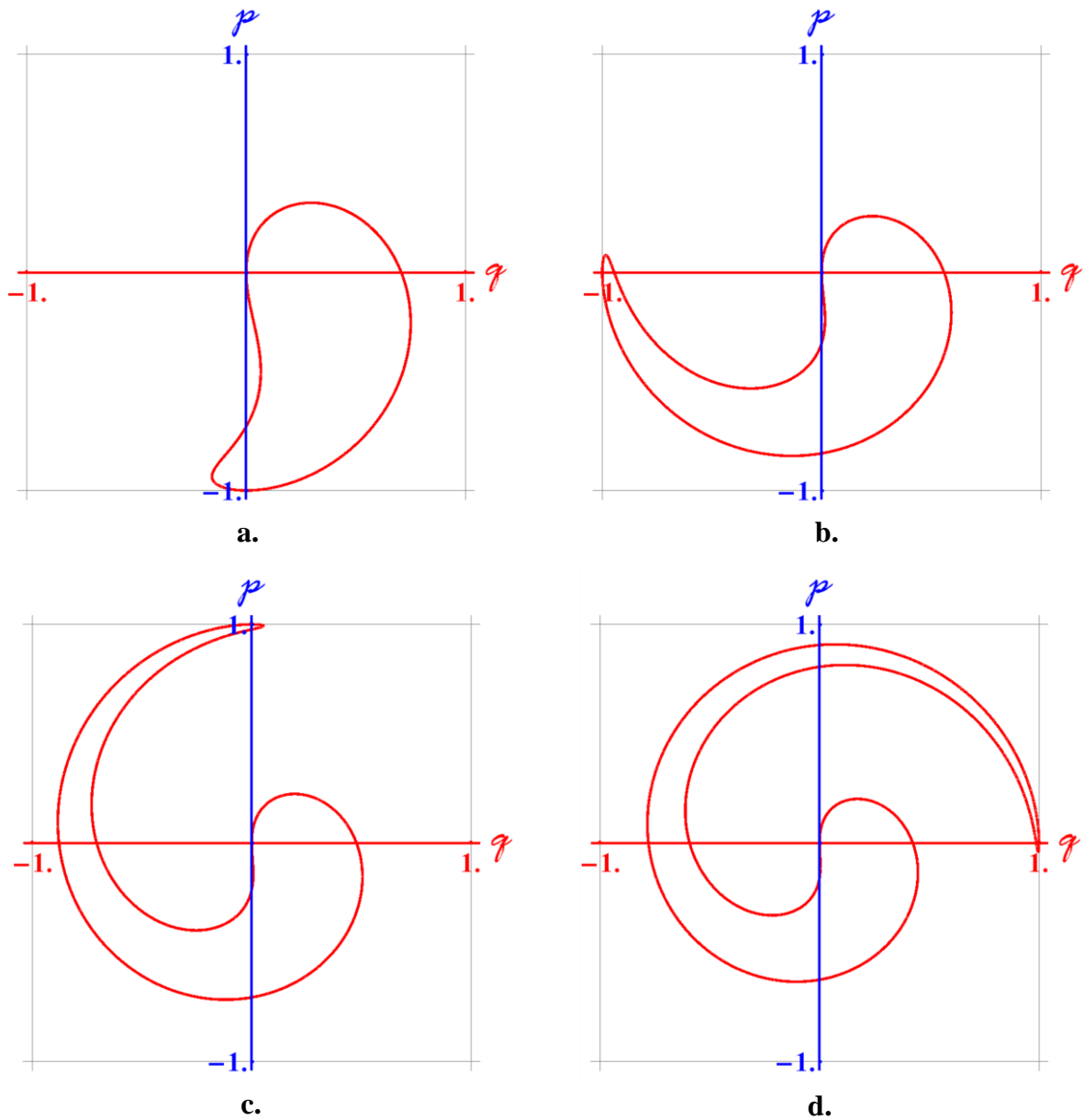

**Fig. (1):** The 2-D time-evolution contours of the classical probability distribution function $Q$ for the anharmonic oscillator in phase space for different values of time ($\tau$): (a) $\tau = \pi/2$, (b) $\tau = \pi$, (c) $\tau = 3\pi/2$, and (d) $\tau = 2\pi$.

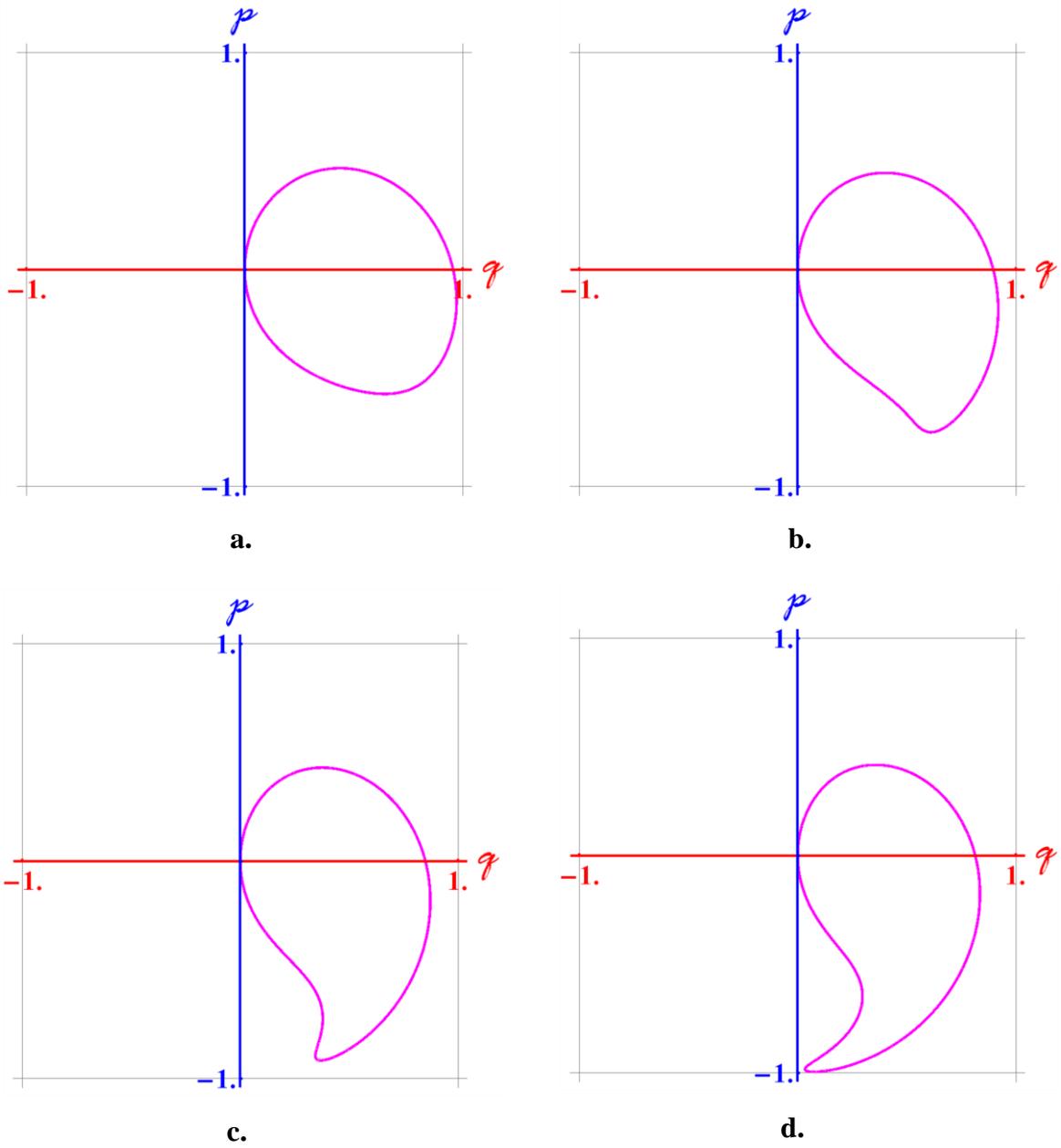

**Fig. (2):** The 2-D time-evolution contours of the classical probability distribution function $\mathcal{P}_{CL}^{q}(\alpha,\alpha^{*};t)$ for the q-deformed harmonic oscillator with frequency $\omega_{q}^{(1)}$ given by eqn. (18a) and $q = 0.5$ in phase space for different values of time ($\tau$): (a) $\tau = \pi/2$, (b) $\tau = \pi$, (c) $\tau = 3\pi/2$, and (d) $\tau = 2\pi$.

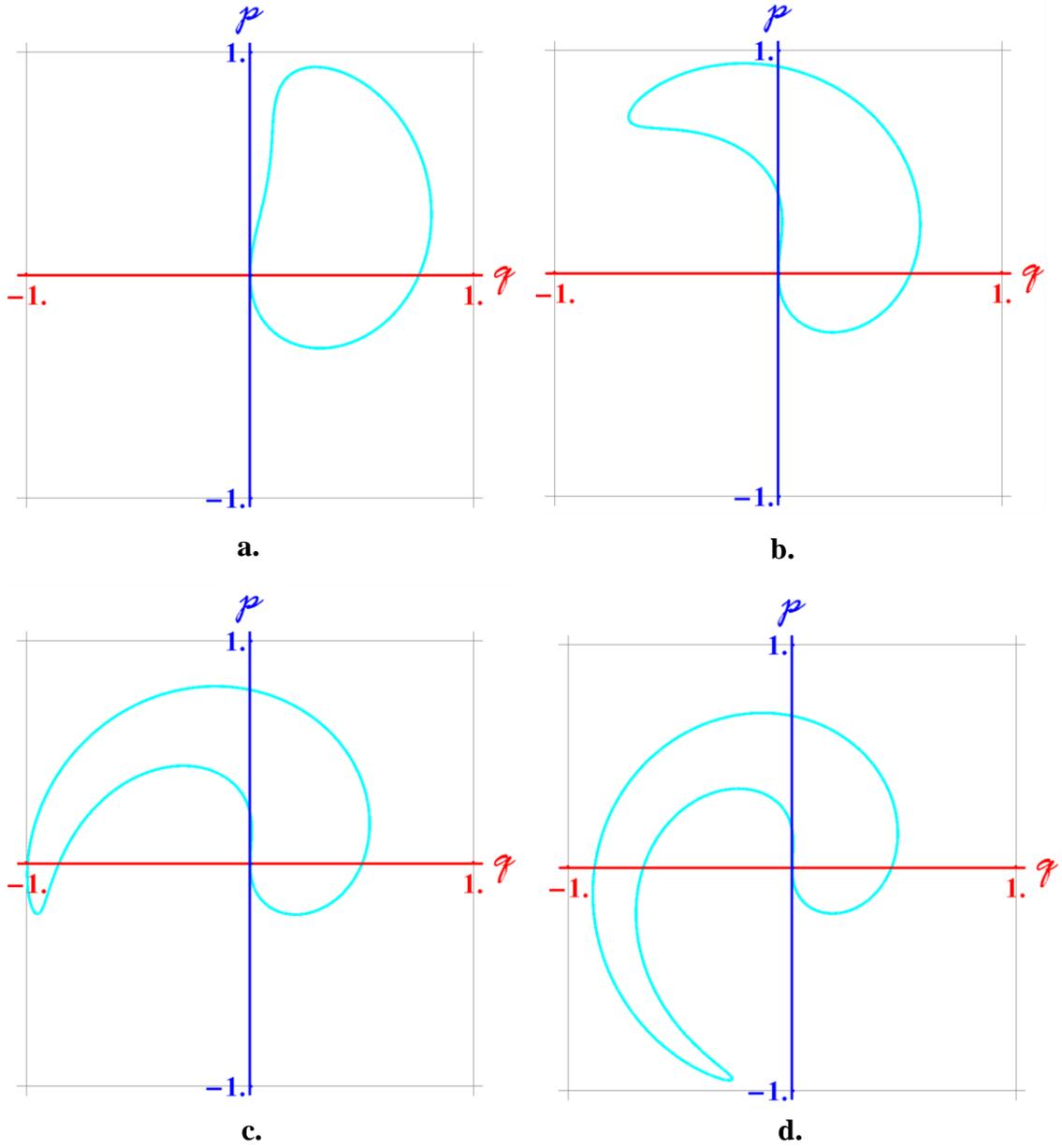

**Fig. (3):** The 2-D time-evolution contours of the classical probability distribution function $\mathcal{P}_{CL}^{q}(\alpha,\alpha^*;t)$ for the q-deformed harmonic oscillator with frequency $\omega_q^{(2)}$ given by eqn. (18b) and $q = 0.5$ in phase space for different values of time ($\tau$): (a) $\tau = \pi/2$, (b) $\tau = \pi$, (c) $\tau = 3\pi/2$, and (d) $\tau = 2\pi$.

In Figs. (4)-(6), results of 3-D time-evolution of the same probability distributions are presented. Eqns. (53) and (54) were used to calculate the values of the probability distributions $\mathcal{P}_{CL}^{q}(\alpha,\alpha^{*};t)$ and $P_{CL}^{q}(\alpha_q,\alpha_q^{*};t)$. It is clear from all of these figures that the peaks of the q-deformed Gaussians for the probability distributions $Q$ and $\mathcal{P}_{CL}^{q}(\alpha,\alpha^{*};t)$ do not change with time and equal to the maximum value (i.e., 1). These peaks follow the trajectories shown in Fig. (1) and Fig. (2) for the classical q-deformed oscillator. Another noticeable feature is the observation that the Gaussian shapes of these distributions become more convoluted around themselves as $t \to \infty$, which is clear in Figs. (4) - (6).

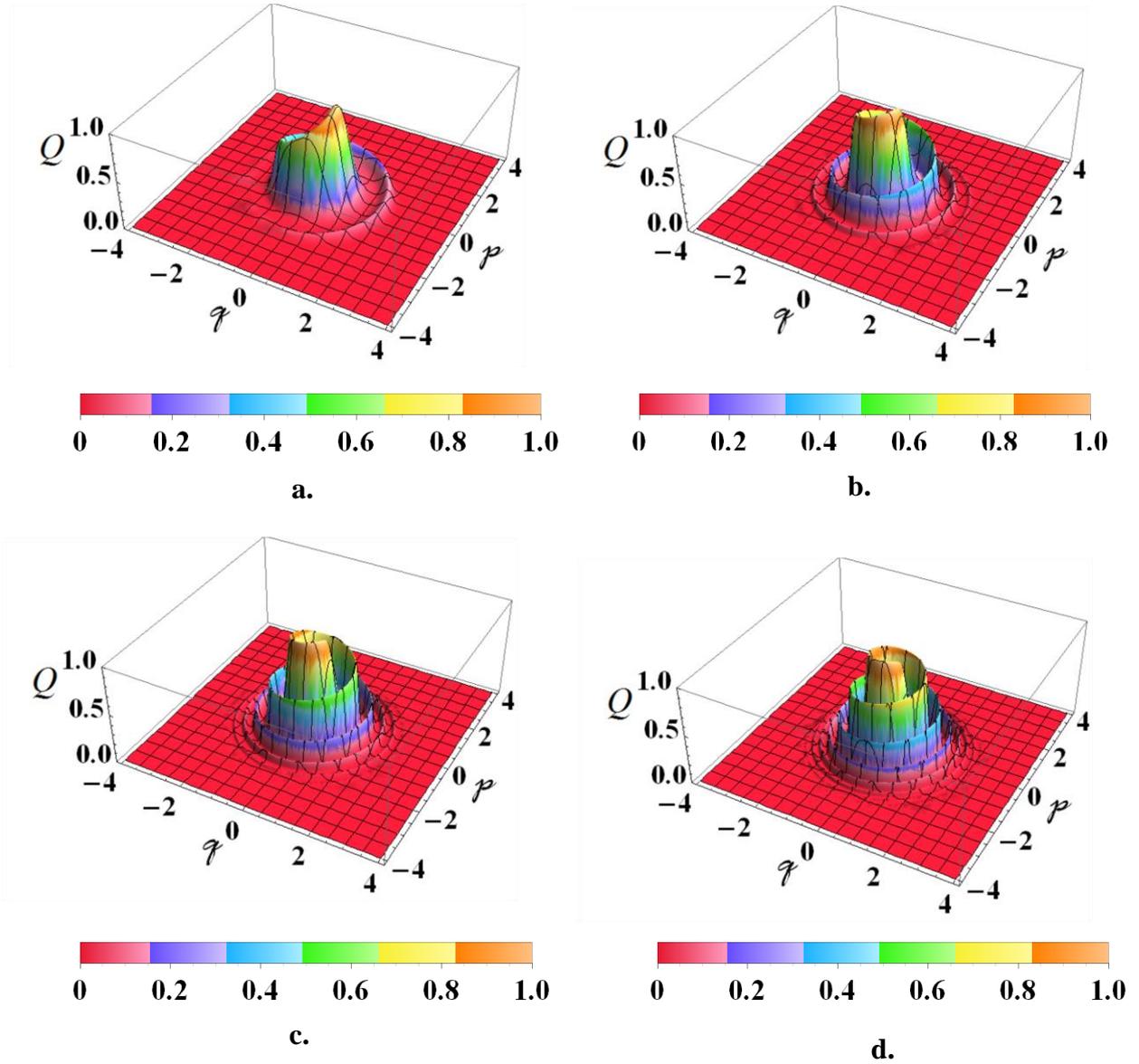

**Fig. (4):** The 3-D time-evolution of the classical probability distribution function $Q$ for the anharmonic oscillator in phase space for different values of time ($\tau$): (a) $\tau = \pi/2$, (b) $\tau = \pi$, (c) $\tau = 3\pi/2$, and (d) $\tau = 2\pi$.

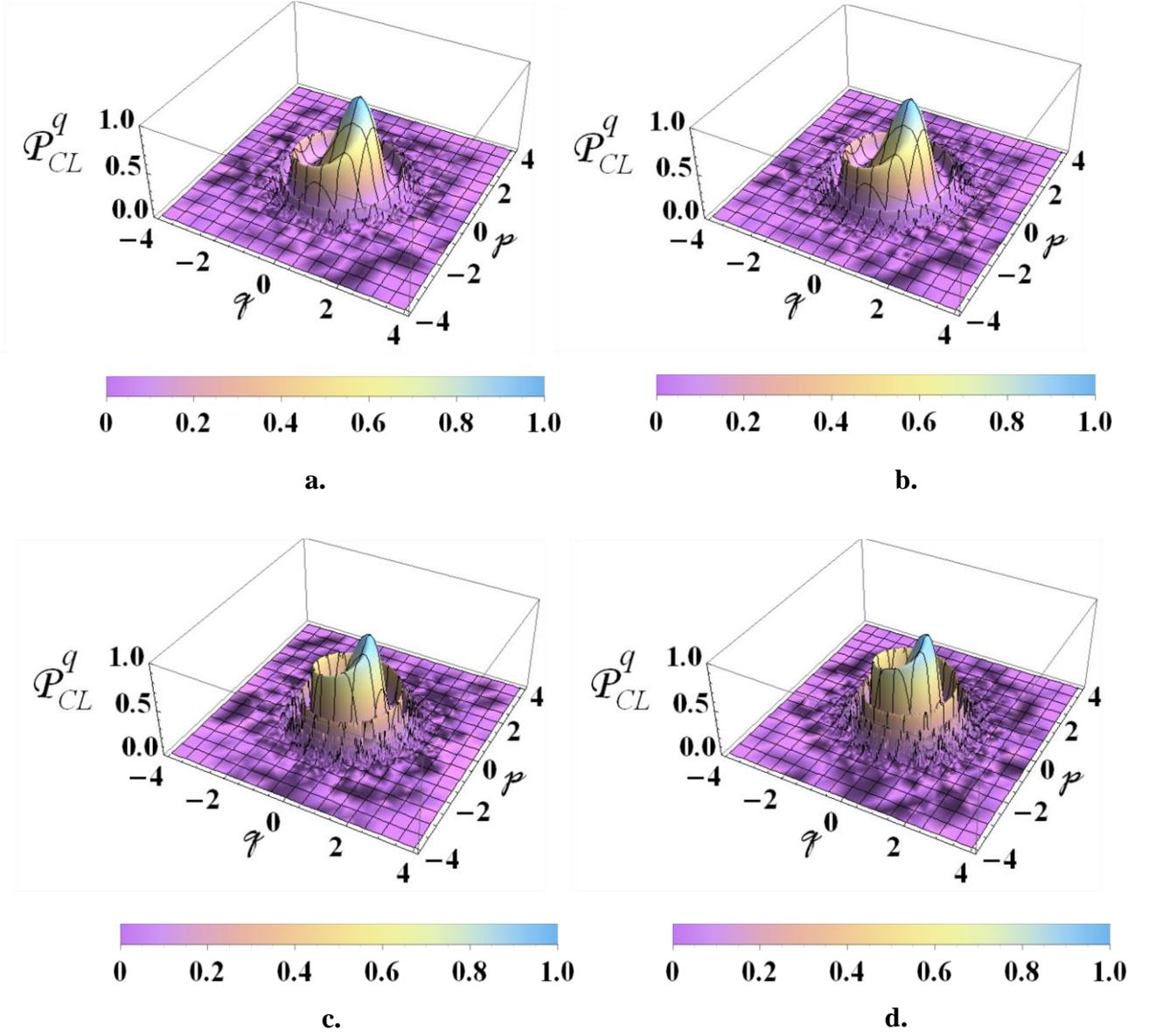

**Fig. (5):** The 3-D time-evolution of the classical probability distribution function $\mathcal{P}_{CL}^{q}(\alpha,\alpha^{*};t)$ for the q-deformed harmonic oscillator with frequency $\omega_{q}^{(1)}$ given by eqn. (18a) and $q=0.5$ in phase space for different values of time ($\tau$): (a) $\tau=\pi/2$, (b) $\tau=\pi$, (c) $\tau=3\pi/2$, and (d) $\tau=2\pi$.

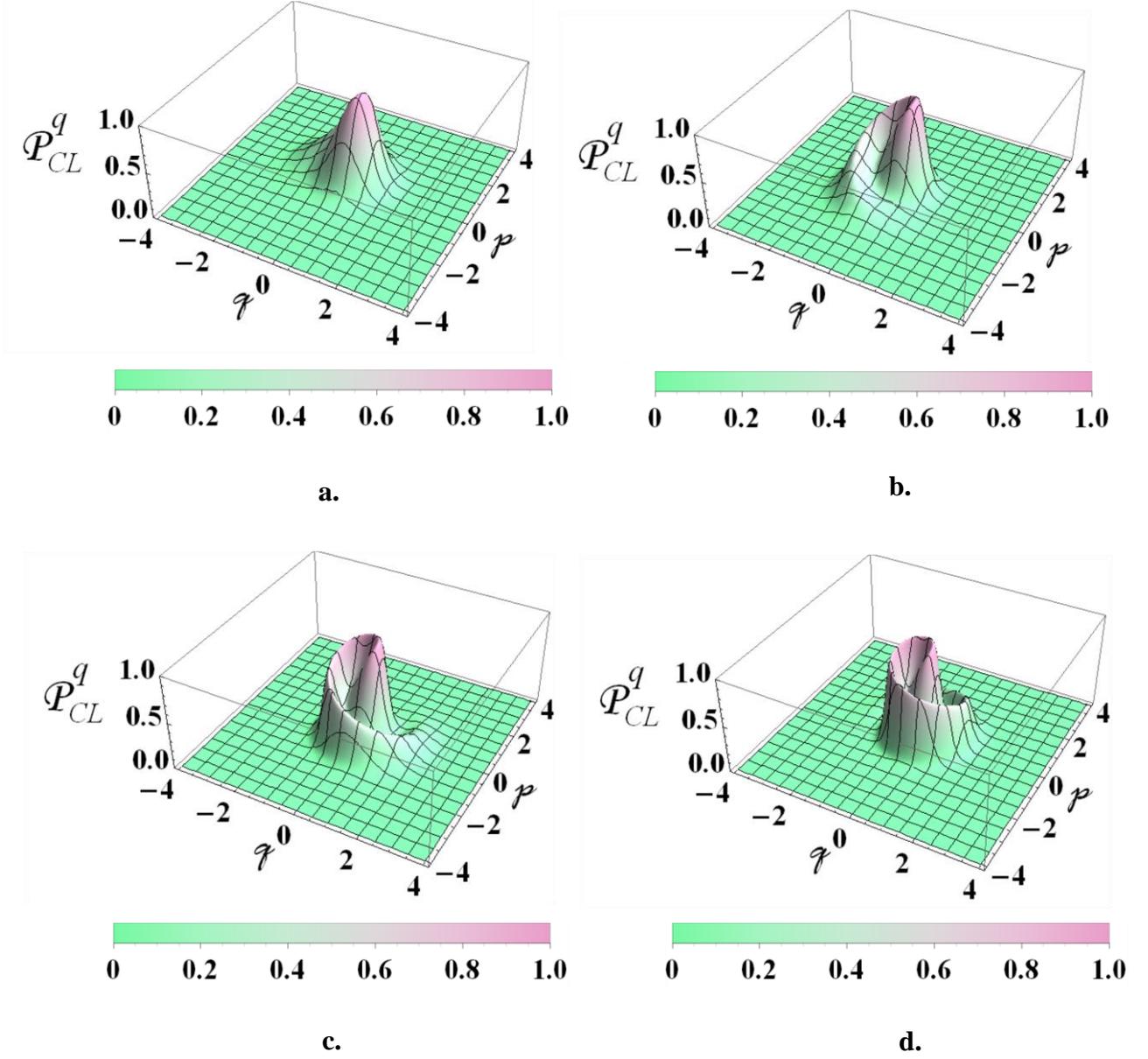

**Fig. (6):** The 3-D time-evolution of the classical probability distribution function $\mathcal{P}_{CL}^{q}(\alpha,\alpha^*;t)$ for the q-deformed harmonic oscillator with frequency $\omega_q^{(2)}$ given by eqn. (18b) and $q = 0.5$ in phase space for different values of time ($\tau$): (a) $\tau = \pi/2$, (b) $\tau = \pi$, (c) $\tau = 3\pi/2$, and (d) $\tau = 2\pi$.

## Conclusions

The investigation of the behavior of the Liouville equation derived in the present work for the q-deformed 1-D classical harmonic oscillator in phase space shows whorl shapes evolving with time as in Fig. (2) and Fig. (3). These figures show many similarities to the results obtained by Milburn [10] for the 1-D classical anharmonic oscillator depicted in Fig. (1). These similarities, and in particular the whorl shapes, result from the fact that the anharmonic oscillator itself represents a kind of deformation with a frequency which is a function of $|\alpha|^2$. This leads us to conclude that the whorl shapes in phase space can be considered as a generalized phenomenon connected with the deformation of any classical system, with q-deformation of the classical harmonic oscillator being a special case.

## Appendices

### Appendix-A

**Evaluation of the Poisson Brackets**

$$\{\alpha, \mathbb{H}_q(\alpha, \alpha^*)\} \text{ and }$$
$$\{\alpha_q, \mathcal{H}_q(\alpha_q, \alpha_q^*)\}$$

According to the definition of the Poisson bracket [12], and considering $\mathbb{H}_q$ as a function of the two independent variables $\alpha$ and $\alpha^*$, one can write:

$$\left(\frac{\partial \mathbb{H}_q}{\partial p}\right)_q = \left(\frac{\partial \mathbb{H}_q}{\partial \alpha}\right)_{\alpha^*} \left(\frac{\partial \alpha}{\partial p}\right)_q + \left(\frac{\partial \mathbb{H}_q}{\partial \alpha^*}\right)_{\alpha} \left(\frac{\partial \alpha^*}{\partial p}\right)_q \quad (A.1)$$

and,

$$\left(\frac{\partial \mathbb{H}_q}{\partial q}\right)_p = \left(\frac{\partial \mathbb{H}_q}{\partial \alpha}\right)_{\alpha^*} \left(\frac{\partial \alpha}{\partial q}\right)_p + \left(\frac{\partial \mathbb{H}_q}{\partial \alpha^*}\right)_{\alpha} \left(\frac{\partial \alpha^*}{\partial q}\right)_p \quad (A.2)$$

Hence,

$$\{\alpha, \mathbb{H}_q(\alpha, \alpha^*)\} =$$
$$\left(\frac{\partial \alpha}{\partial q}\right)_p \left(\left(\frac{\partial \mathbb{H}_q}{\partial \alpha}\right)_{\alpha^*} \left(\frac{\partial \alpha}{\partial p}\right)_q + \left(\frac{\partial \mathbb{H}_q}{\partial \alpha^*}\right)_{\alpha} \left(\frac{\partial \alpha^*}{\partial p}\right)_q\right)$$
$$- \left(\left(\frac{\partial \alpha}{\partial p}\right)_q \left(\frac{\partial \mathbb{H}_q}{\partial \alpha}\right)_{\alpha^*} \left(\frac{\partial \alpha}{\partial q}\right)_p + \left(\frac{\partial \mathbb{H}_q}{\partial \alpha^*}\right)_{\alpha} \left(\frac{\partial \alpha^*}{\partial q}\right)_p\right) \quad (A.3)$$

Eqn. (A.3) can be simplified to obtain:

$$\{\alpha, \mathbb{H}_q(\alpha, \alpha^*)\} =$$
$$\left(\left(\frac{\partial \alpha}{\partial q}\right)_p \left(\frac{\partial \alpha^*}{\partial p}\right)_q - \left(\frac{\partial \alpha}{\partial p}\right)_q \left(\frac{\partial \alpha^*}{\partial q}\right)_p\right) \left(\frac{\partial \mathbb{H}_q}{\partial \alpha^*}\right)_{\alpha} \quad (A.4)$$

Now, since

$$\{\alpha, \mathbb{H}_q(\alpha, \alpha^*)\}_{\alpha, \alpha^*} = \left(\frac{\partial \mathbb{H}_q}{\partial \alpha^*}\right)_{\alpha} \quad (A.5)$$

then, substituting eqn. (A.5) into eqn. (A.4), one obtains:

$$\{\alpha, \mathbb{H}_q(\alpha, \alpha^*)\} =$$
$$\{\alpha, \alpha^*\} \cdot \{\alpha, \mathbb{H}_q(\alpha, \alpha^*)\}_{\alpha, \alpha^*} \quad (A.6)$$

Similarly, one can prove that

$$\{\alpha_q, \mathcal{H}_q(\alpha_q, \alpha_q^*)\} =$$
$$\{\alpha_q, \alpha_q^*\} \cdot \{\alpha_q, \mathcal{H}_q(\alpha_q, \alpha_q^*)\}_{\alpha_q, \alpha_q^*} \quad (A.7)$$

**Appendix-B**

**Evaluation of the Poisson Brackets**
$\{\mathcal{H}_q(\alpha_q, \alpha_q^*), P_{CL}^q(\alpha_q, \alpha_q^*; t)\}$ **and**
$\{\mathbb{H}_q(\alpha, \alpha^*), \mathcal{P}_{CL}^q(\alpha, \alpha^*; t)\}$

Considering $\mathcal{H}_q$ as a function of the two independent variables $\alpha_q$ and $\alpha_q^*$, one can write:

$$\left(\frac{\partial \mathcal{H}_q}{\partial q}\right)_p = \left(\frac{\partial \mathcal{H}_q}{\partial \alpha_q}\right)_{\alpha_q^*} \left(\frac{\partial \alpha_q}{\partial q}\right)_p$$
$$+ \left(\frac{\partial \mathcal{H}_q}{\partial \alpha_q^*}\right)_{\alpha_q} \left(\frac{\partial \alpha_q^*}{\partial q}\right)_p \quad (B.1)$$

and,

$$\left(\frac{\partial \mathcal{H}_q}{\partial p}\right)_q = \left(\frac{\partial \mathcal{H}_q}{\partial \alpha_q}\right)_{\alpha_q^*} \left(\frac{\partial \alpha_q}{\partial p}\right)_q$$
$$+ \left(\frac{\partial \mathcal{H}_q}{\partial \alpha_q^*}\right)_{\alpha_q} \left(\frac{\partial \alpha_q^*}{\partial p}\right)_q \quad (B.2)$$

Substituting eqns. (B.1) and (B.2) into the definition of the Poisson bracket $\{\mathcal{H}_q(\alpha_q, \alpha_q^*), P_{CL}^q(\alpha_q, \alpha_q^*; t)\}$, the result becomes:

$$\{\mathcal{H}_q(\alpha_q, \alpha_q^*), P_{CL}^q(\alpha_q, \alpha_q^*; t)\} =$$

$$\left(\left(\frac{\partial \mathcal{H}_q}{\partial \alpha_q}\right)_{\alpha_q^*} \left(\frac{\partial \alpha_q}{\partial q}\right)_p\right.$$
$$\left. + \left(\frac{\partial \mathcal{H}_q}{\partial \alpha_q^*}\right)_{\alpha_q} \left(\frac{\partial \alpha_q^*}{\partial q}\right)_p\right)$$

$$\cdot \left(\left(\frac{\partial P_{CL}^q}{\partial \alpha_q}\right)_{\alpha_q^*} \left(\frac{\partial \alpha_q}{\partial p}\right)_q\right.$$
$$\left. + \left(\frac{\partial P_{CL}^q}{\partial \alpha_q^*}\right)_{\alpha_q} \left(\frac{\partial \alpha_q^*}{\partial p}\right)_q\right)$$

$$- \left(\left(\frac{\partial \mathcal{H}_q}{\partial \alpha_q}\right)_{\alpha_q^*} \left(\frac{\partial \alpha_q}{\partial p}\right)_q\right.$$
$$\left. + \left(\frac{\partial \mathcal{H}_q}{\partial \alpha_q^*}\right)_{\alpha_q} \left(\frac{\partial \alpha_q^*}{\partial p}\right)_q\right)$$

$$\cdot \left(\left(\frac{\partial P_{CL}^q}{\partial \alpha_q}\right)_{\alpha_q^*} \left(\frac{\partial \alpha_q}{\partial q}\right)_p\right.$$
$$\left. + \left(\frac{\partial P_{CL}^q}{\partial \alpha_q^*}\right)_{\alpha_q} \left(\frac{\partial \alpha_q^*}{\partial q}\right)_p\right) \quad (B.3)$$

After some mathematical manipulations, this equation can be simplified to yield:

$$\{\mathcal{H}_q(\alpha_q,\alpha_q^*), P_{CL}^q(\alpha_q,\alpha_q^*;t)\} =$$

$$\left(\frac{\partial \mathcal{H}_q}{\partial \alpha_q^*}\right)_{\alpha_q} \left(\frac{\partial P_{CL}^q}{\partial \alpha_q}\right)_{\alpha_q^*}$$

$$\cdot \left(\left(\frac{\partial \alpha_q^*}{\partial q}\right)_p \left(\frac{\partial \alpha_q}{\partial p}\right)_q \right.$$

$$\left. - \left(\frac{\partial \alpha_q^*}{\partial p}\right)_q \left(\frac{\partial \alpha_q}{\partial q}\right)_p\right)$$

$$+ \left(\frac{\partial \mathcal{H}_q}{\partial \alpha_q}\right)_{\alpha_q^*} \left(\frac{\partial P_{CL}^q}{\partial \alpha_q^*}\right)_{\alpha_q}$$

$$\cdot \left(\left(\frac{\partial \alpha_q}{\partial q}\right)_p \left(\frac{\partial \alpha_q^*}{\partial p}\right)_q \right.$$

$$\left. - \left(\frac{\partial \alpha_q}{\partial p}\right)_q \left(\frac{\partial \alpha_q^*}{\partial q}\right)_p\right)$$

(B.4)

But since the Poisson bracket $\{\alpha_q, \alpha_q^*\} =$

$$\left(\frac{\partial \alpha_q}{\partial q}\right)_p \left(\frac{\partial \alpha_q^*}{\partial p}\right)_q - \left(\frac{\partial \alpha_q}{\partial p}\right)_q \left(\frac{\partial \alpha_q^*}{\partial q}\right)_p$$

(B.5)

then, substituting eqn. (B.5) into eqn. (B.4) one gets:

$$\{\mathcal{H}_q(\alpha_q,\alpha_q^*), P_{CL}^q(\alpha_q,\alpha_q^*;t)\} = \{\alpha_q,\alpha_q^*\}$$

$$\cdot \{\mathcal{H}_q(\alpha_q,\alpha_q^*), P_{CL}^q(\alpha_q,\alpha_q^*;t)\}_{\alpha_q,\alpha_q^*}$$

(B.6)

Similarly, one can prove that

$$\{\mathbb{H}_q(\alpha,\alpha^*), \mathcal{P}_{CL}^q(\alpha,\alpha^*;t)\} = \{\alpha,\alpha^*\}$$

$$\cdot \{\mathbb{H}_q(\alpha,\alpha^*), \mathcal{P}_{CL}^q(\alpha,\alpha^*;t)\}_{\alpha,\alpha^*}$$

(B.7)